\documentclass[a4paper,showpacs]{iopart}
\pdfoutput=1 
\usepackage[utf8]{inputenc}
\usepackage[T1]{fontenc}
\usepackage[usenames,dvipsnames]{xcolor}
\usepackage{graphicx}
\usepackage{iopams}
\usepackage{setstack}
\usepackage{amsbsy}

\usepackage[]{todonotes}

\definecolor{darkblue}{rgb}{0,0,0.6}
\definecolor{darkred}{rgb}{0.6,0,0}

\usepackage[colorlinks=true,urlcolor=darkblue,citecolor=darkblue,linkcolor=darkred,hyperfootnotes=false]{hyperref}

\newcommand{\ind}[1]{_\mathrm{#1}}

\renewcommand{\emph}[1]{\textit{#1}}

\newcommand{\transp}{^\mathrm{T}}

\newcommand{\dd}{{\mathrm{d}}}

\newcommand{\zero}{{\boldsymbol{0}}}
\newcommand{\un}{{\boldsymbol{1}}}

\newcommand{\ff}{{\boldsymbol{f}}}

\newcommand{\kk}{{\boldsymbol{k}}}
\newcommand{\pp}{{\boldsymbol{p}}}

\newcommand{\xx}{{\boldsymbol{x}}}

\newcommand{\FF}{{\boldsymbol{F}}}
\newcommand{\GG}{{\boldsymbol{G}}}
\newcommand{\xxi}{{\boldsymbol{\xi}}}
\newcommand{\XXi}{{\boldsymbol{\Xi}}}
\newcommand{\eeta}{{\boldsymbol{\eta}}}
\newcommand{\nnabla}{{\boldsymbol{\nabla}}}

\begin{document}

\title[Generalized Langevin equations for a driven tracer in dense soft colloids]{Generalized Langevin equations for a driven tracer in dense soft colloids: construction and applications}

\author{Vincent D\' emery$^{1,2}$}
\address{$^1$ Laboratoire de Physique Th\'eorique de la Mati\`ere Condens\'ee, CNRS/UPMC,\\ 4 Place Jussieu, 75005 Paris, France.}
\address{$^2$ Department of Physics, University of Massachusetts, Amherst, MA 01003, USA.}
\ead{vdemery@physics.umass.edu}

\author{Olivier B\' enichou}
\address{Laboratoire de Physique Th\'eorique de la Mati\`ere Condens\'ee, CNRS/UPMC,\\ 4 Place Jussieu, 75005 Paris, France.}

\author{Hugo Jacquin}
\address{Universit\'e de Lyon, Laboratoire de Physique, \'Ecole Normale Sup\'erieure de Lyon, CNRS,
46 all\'ee d'Italie, F-69007 Lyon, France}

\begin{abstract}
We describe a tracer in a bath of soft Brownian colloids by a particle coupled to the density field of the other bath particles. From the Dean equation, we derive an exact equation for the evolution of the whole system, and show that the density field evolution can be linearized in the limit of a dense bath. 
This linearized Dean equation with a tracer taken apart is validated by the reproduction of previous results on the mean-field liquid structure and transport properties. 
Then, the tracer is submitted to an external force and we compute the density profile around it, its mobility and its diffusion coefficient. 
Our results exhibit effects such as bias enhanced diffusion that are very similar to those observed in the opposite limit of a hard core lattice gas, indicating the robustness of these effects.
Our predictions are successfully tested against molecular dynamics simulations.
\end{abstract}


\maketitle

\section{Introduction}\label{}

Transport in a crowded environment is an issue particularly relevant to cell biology, where the crowding inside the cytoplasm can strongly affect the molecular diffusion~\cite{Bressloff2013, Hofling2013, Tabei2013, Szymanski2009} and thus hinder reactivity.
This effect is also present in the plasma membrane, where the high protein concentration may slow down diffusion~\cite{Ramadurai2009}. 
The transport properties of these media can be measured globally, using for example fluorescence recovery after photobleaching~\cite{Konopka2006} or fluorescence correlation microscopy~\cite{Szymanski2009}, or locally, with single particle tracking~\cite{Tabei2013}.

Conversely, the observation of the motion of a probe is used in microrheology to investigate the properties of such complex fluids~\cite{Waigh2005, Wilson2011b}. 
In \emph{passive} microrheology, the probe diffuses freely or oscillates in the parabolic well created by optical or magnetic tweezers and the measurement of its diffusion coefficient is used to determine, via the Stokes-Einstein relation, the solvent viscosity~\cite{Wilson2009}. The probe can also be pulled by tweezers and the drag force is measured: this is \emph{active} microrheology.
However, when the size of the probe becomes comparable to the size of the complex fluid constituents, its motion is no longer a pure Brownian motion and is thus harder to analyse. This is one of the reasons why there is no straightforward relation between microrheology and macrorheology measurements~\cite{Wilson2009}.
Some theoretical studies have addressed the motion of the probe in complex fluids~\cite{Squires2005, Kruger2009, Squires2010}, but they are limited to an hydrodynamic description of the complex fluid or to dilute colloidal suspensions. 

It has been proposed to model dense colloidal assemblies by a gas of hard core particles on a lattice~\cite{Benichou2000, Benichou2013b, Benichou2013d, Benichou2013e}. These studies focused on the motion of a tracer submitted to an external bias and computed its diffusion coefficient and the probability density function of its position. They exhibited unexpected effects such as bias enhanced diffusion coefficient or even super-diffusion in a very dense environment.

\begin{figure}[b]
\begin{center}
\includegraphics[width=0.7\linewidth]{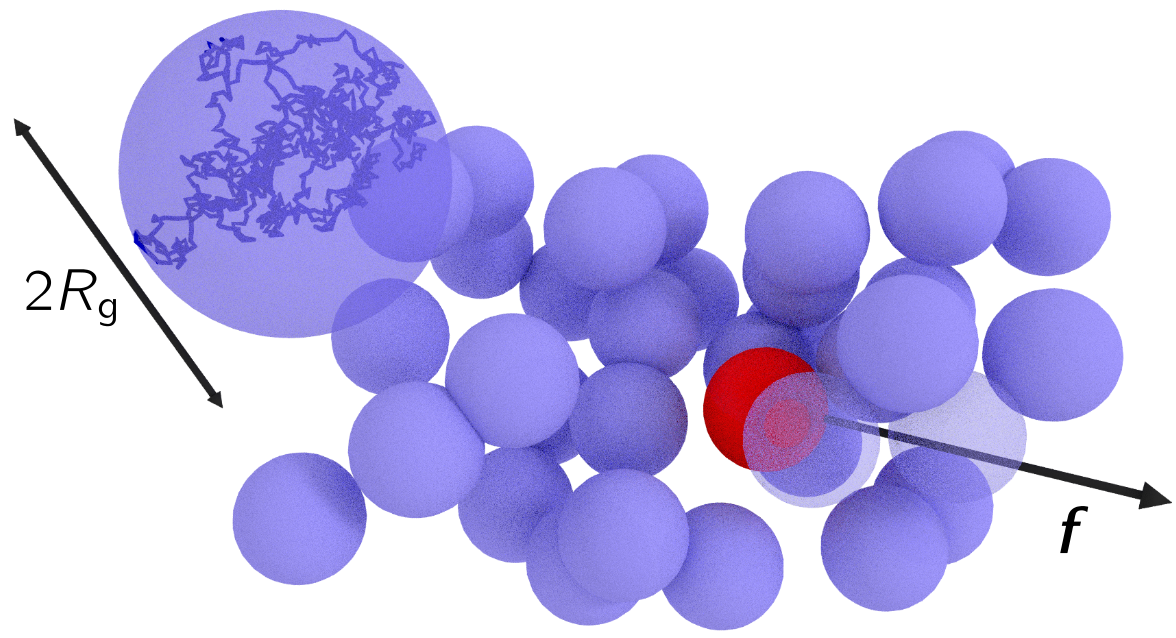}
\end{center}
\caption{(Colour online) 
Colloidal solution where a tracer (red particle) is pulled by an external force $\ff$. The soft colloids may represent polymer coils, the size of the colloids being the gyration radius $R\ind{g}$ of the polymer (top left).
}
\label{fig_schema}
\end{figure}

Here, we consider the opposite limit of soft colloids, where we address the effective mobility and diffusion coefficient of a tracer submitted to an external force (cf. Figure~\ref{fig_schema}). Soft colloids can be polymer coils, that interact with an effective potential that is close to a Gaussian~\cite{Louis2000b} or disordered proteins, such as the $\alpha$-synuclein~\cite{Wang2012}. Colloids or macromolecules motion in a solvent is well captured by an overdamped Langevin equation~\cite{Hofling2013}. Transport in such system has been addressed by various tools, such as mode-coupling theory~\cite{Kob2003, Berthier2010} or direct perturbative analysis~\cite{Dean2004}. 
In this work, we pursue a perturbative treatment of the intermolecular interaction, that is relevant when the molecules are soft, i.e. when the pair interaction is weak.

The overall density of a system of Langevin particles evolves according to the Dean equation~\cite{Dean1996}.
In order to resolve the tracer dynamics, we treat it separately and gather the other particles in a partial density field: two coupled evolution equations rule the whole system. 
At high density, we show that the density field evolution can be approximated by a much simpler linear equation.
The final set of equations allows to reproduce easily former results such as the pair correlation function in liquids under the mean-field approximation~\cite{Likos2001} or the effective diffusion coefficient of the tracer without any external forcing~\cite{Dean2004}.
Then we investigate the effect of an external forcing on the tracer and get analytical expressions for the average density around the tracer and the tracer effective mobility and diffusion coefficient. 
We show that they are qualitatively strikingly close to those computed for a tracer pulled in a hard core lattice gas. Notably, we find a critical force above which the diffusion is enhanced and the same decaying exponent for the density perturbation behind the tracer.

This article is organized as follows. The model is defined with the observables we focus on in section~\ref{sec_model}. The linearized Dean equation with the tracer taken apart~(LDT) is derived in section~\ref{sec_lin_dean}. The LDT is applied to a tracer in the absence of an external force in section~\ref{sec:tracer_no_force}\ and previous results are recovered. 
In section~\ref{sec_mob_diff}, we apply the LDT to a tracer submitted to an external force and compute the density profile around it, its effective mobility and its effective diffusion coefficient. 
These results are compared to molecular dynamics simulations in section~\ref{sec:numerics}. We conclude in section~\ref{sec_conclu}.

\section{Model}\label{sec_model}

We consider $N+1$ Brownian particles interacting via the pair potential $V(\xx)$ and located at $\xx_i$ in a $d$-dimensional space; 
the unit of length is the particle size and the unit of energy is the characteristic energy of the interaction.
Moreover, an external force $\ff$ is applied to the tracer, identified by $i=0$.
The motion of each particle follows an overdamped Langevin dynamics:
\begin{equation}\label{eq:langevin_npart}
\dot \xx_i(t)=\delta_{i,0}\ff-\sum_{j\neq i} \nnabla_{\xx_i} V(\xx_i(t)-\xx_j(t))+\eeta_i(t).
\end{equation}
where $\eeta_i(t)$ is a Gaussian white noise with correlation function
\begin{equation}\label{eq:noise_part}
\left\langle \eeta_i(t)\eeta_j(t')\transp \right\rangle=2T\delta_{i,j}\delta(t-t')\un
\end{equation}
and $T$ stands for the thermal energy in units of the pair potential characteristic energy.

First, we are interested in the density field of all the particles but the tracer, that is defined by
\begin{equation}
\rho(\xx,t)=\sum_{i=1}^N \delta(\xx-\xx_i(t)).
\end{equation}
Note that the sum does not include the tracer. More precisely, we look at the average density field in the reference frame of the tracer:
\begin{equation}\label{eq:av_dens_tracer}
\langle \rho^*(\xx) \rangle=\langle \rho(\xx+\xx_0(t),t) \rangle.
\end{equation}
The superscript ${}^*$ represents the reference frame of the tracer and the time dependence has been removed since we focus on the stationary state.
We define the origin of the coordinate system so that
\begin{equation}
\xx_0(t=0)=\zero.
\end{equation}
Then, we want to compute two observables describing the particle dynamical properties. 
The first observable we are interested in is the tracer effective mobility $\kappa\ind{eff}$, defined by
\begin{equation}\label{eq:def_mobility}
\langle\xx_0(t)\rangle\underset{t\rightarrow\infty}{\sim} \kappa\ind{eff} \ff t,
\end{equation}
and the second is its effective diffusion coefficient $D\ind{eff}$, defined by
\begin{equation}
\left\langle [\xx_0(t) - \langle \xx_0(t) \rangle]^2 \right\rangle\underset{t\rightarrow\infty}{\sim}2dD\ind{eff} t.
\end{equation}
In the absence of interaction with the other particles, the tracer undergoes a biased Brownian motion of bare mobility $\kappa_\xx=1$ and bare diffusion coefficient $D_\xx=T$, independently of the bias.

\section{Linearized Dean equation}\label{sec_lin_dean}

\subsection{Derivation}\label{}

We want to describe the evolution of the $N+1$ particles as the evolution of the tracer $\xx_0(t)$ coupled to the density field $\rho(\xx,t)$ of the other particles.
The Dean equation~\cite{Dean1996} gives the evolution of the total density $\rho\ind{tot}(\xx,t)=\delta(\xx-\xx_0(t))+\rho(\xx,t)$, but it appears in its proof that the tracer can be extracted to get the evolution equation of the partial density $\rho(\xx,t)$ (see \ref{ap_dean_1p} for more details):
\begin{equation}\label{eq:dean}
\partial_t\rho=T\nnabla^2\rho + \nnabla\cdot[\rho \nnabla (V*\rho\ind{tot})]+\nnabla\cdot \left(\rho^{1/2} \xxi\right);
\end{equation}
the star $*$ denotes the convolution and $\xxi$ is a Gaussian white noise with correlation function
\begin{equation}\label{eq:noise_dens}
\left\langle \xxi(\xx,t)\xxi(\xx',t')\transp \right\rangle=2T\delta(t-t')\delta(\xx-\xx')\un.
\end{equation}
The Langevin equation (\ref{eq:langevin_npart}) for the tracer can be rewritten with the density field $\rho(\xx,t)$:
\begin{equation}\label{eq:langevin_partindens}
\dot \xx_0(t)=\ff-\nnabla (V*\rho)(\xx_0(t),t)+\eeta(t).
\end{equation}
The noise is the one appearing in Eq.~(\ref{eq:langevin_npart}), $\eeta(t)=\eeta_0(t)$.
Equations (\ref{eq:dean}-\ref{eq:langevin_partindens}) are exact. 
However, the density evolution~(\ref{eq:dean}) is non linear and contains a multiplicative noise;
we show that it can be linearized if the bath is dense.

We write the density created by the $N$ particles as the sum of a constant uniform term and a fluctuating term:
\begin{equation}
\rho(\xx,t)=\rho_0+\rho_0^{1/2}\phi(\xx,t).
\end{equation}
The uniform density is $\rho_0=N/\mathcal{V}$, $\mathcal{V}$ being the volume of the system. Our computations are done in the limit of an infinite system size, with $\rho_0$ kept constant.
We also define the rescaled interaction potential as
\begin{equation}
v(\xx)=\rho_0 V(\xx).
\end{equation}
The evolution of the density fluctuations $\phi(\xx,t)$ reads
\begin{eqnarray}\label{eq:evol_phi_comp}
\fl\partial_t\phi=T\nnabla^2\phi+\nnabla^2(v*\phi) + \rho_0^{-1/2}\nnabla\cdot[\phi\nnabla(v*\phi)] +  \rho_0^{-1/2}\nnabla^2(v*\delta_{\xx_0}) \nonumber\\
 + \rho_0^{-1}\nnabla\cdot[\phi\nnabla (v*\delta_{\xx_0})] +\nnabla\cdot\left[\left(1+\rho_0^{-1/2}\phi\right)^{1/2}\xxi\right],
\end{eqnarray}
where $\delta_{\xx_0}$ is the Dirac delta function centered at $\xx_0$.

The quadratic terms in the field $\phi$ and the multiplicative noise are negligible when $\rho_0^{-1/2}\phi\ll 1$. In this case the equation for the density fluctuations reduces to
\begin{equation}\label{eq:evol_phi_lin}
\partial_t\phi=T\nnabla^2\phi+\nnabla^2(v*\phi) + \rho_0^{-1/2}\nnabla^2(v*\delta_{\xx_0}) +\nnabla\cdot\xxi.
\end{equation}
The density deviations from the average density $\rho_0$ come from the presence of the tracer and from the thermal fluctuations (represented respectively by the third and fourth terms in (\ref{eq:evol_phi_lin})).
From this linear equation, these contributions can be evaluated: the effect of the tracer on the density fluctuations is of order $\phi\ind{tr}\sim \rho_0^{1/2}V/(T+\rho_0 V)$ and the thermal fluctuations are of order $\phi\ind{th}\sim (1+\rho_0 V/T)^{-1/2}$.
Hence, we can deduce that the condition $\rho_0^{-1/2}\phi\ll 1$ is satisfied in the high density limit
\begin{equation}\label{eq:valid_ldt}
\rho_0\gg 1.
\end{equation}
The tracer equation of motion is given by (\ref{eq:langevin_partindens}):
\begin{equation}\label{eq:langevin_partinphi}
\dot \xx_0(t)=\ff-\rho_0^{-1/2}\nnabla (v*\phi )(\xx_0(t),t)+\eeta(t).
\end{equation}
The set of equations (\ref{eq:evol_phi_lin},\ref{eq:langevin_partinphi}), with the noises correlation functions (\ref{eq:noise_part}, \ref{eq:noise_dens}), forms the linearized Dean equation with a tracer (LDT) and is a first result of our approach. We repeat these equations below for clarity:
\begin{eqnarray}
\dot \xx_0(t) & =\ff-\rho_0^{-1/2}\nnabla (v*\phi )(\xx_0(t),t)+\eeta(t), \label{eq:ldt1}\\
\partial_t\phi & =T\nnabla^2\phi+\nnabla^2(v*\phi) + \rho_0^{-1/2}\nnabla^2(v*\delta_{\xx_0}) +\nnabla\cdot\xxi,
\end{eqnarray}
where the noises have the following correlation functions
\begin{eqnarray}
\left\langle \eeta(t)\eeta(t')\transp \right\rangle & =2T\delta(t-t')\un, \\
\left\langle \xxi(\xx,t)\xxi(\xx',t')\transp \right\rangle & =2T\delta(t-t')\delta(\xx-\xx')\un.\label{eq:ldt4}
\end{eqnarray}

Note that without the tracer taken apart, Eq.~(\ref{eq:evol_phi_lin}) can be written
\begin{equation}\label{eq:dean_lin}
\partial_t\phi=T\nnabla^2\phi+\nnabla^2(v*\phi) +\nnabla\cdot\xxi.
\end{equation}
This evolution is linear and free of multiplicative noise; we show later that it allows to recover mean-field results for the density two-point correlation function.

\subsection{Link to a more general formalism}\label{}

The situation of a tracer interacting with a fluctuating field is ubiquitous and arises also, for example, when a diffusing membrane protein is coupled to the membrane curvature~\cite{Reister2005, Leitenberger2008, Naji2009, Reister_Gottfried2010}. For this reason, a framework has been developed for this kind of systems~\cite{Demery2010,Demery2011}, allowing to derive very general results for the drag force felt by a tracer pulled at constant velocity~\cite{Demery2010, Demery2010a} or the effective diffusion coefficient of a free tracer~\cite{Demery2011, Dean2011}.
We adopt the following strategy: first, we cast the linearized Dean equation with a tracer in this general formalism. 
This allows us to apply previous results to the tracer in a colloidal bath without an external force, that is done in the next section. 
Second, we extend the general formalism~\cite{Demery2011} to cover the case of a tracer biased by an external force.
Finally, we apply the general results to a tracer pulled in a bath of soft spheres.

The general equations describing the evolution of a particle interacting with a fluctuating field read~\cite{Demery2010,Demery2011}
\begin{eqnarray}
\dot\xx_0(t) & = \ff + h\nnabla K\phi(\xx_0(t),t)+\eeta(t),\label{eq:langevin_part_gen}\\
\partial_t\phi(\xx,t) & = - R\Delta\phi(\xx,t)+hRK\delta_{\xx_0(t)}+\xi(\xx,t), \label{eq:langevin_field_gen}
\end{eqnarray}
where $K$, $R$ and $\Delta$ are functional operators and the Gaussian white noises $\eeta(t)$ and $\xi(\xx,t)$ obey
\begin{eqnarray}
\left\langle \eeta(t)\eeta(t')\transp \right\rangle & =2T\delta(t-t')\un,\\
\left\langle \xi(\xx,t)\xi(\xx',t') \right\rangle & = 2T\delta(t-t') R(x-x').\label{eq:field_noise}
\end{eqnarray}
The notations used for functional operators are defined in \ref{ap:operators}.
The mapping between the general formalism and the LDT (\ref{eq:ldt1}-\ref{eq:ldt4}) is given in Fourier space by:
\begin{eqnarray}
h&=\rho_0^{-1/2},\label{eq:map_coupling}\\
\tilde\Delta(\kk) & = T+\tilde v(\kk),\label{eq:map_delta}\\
\tilde R(\kk) & = \kk^2,\\
\tilde K(\kk) & = -\tilde v(\kk).\label{eq:map_form}
\end{eqnarray}

Here, we perform computations perturbatively in the coupling constant $h$; this is also the case of some former studies~\cite{Demery2011,Dean2011}. It is clear from (\ref{eq:langevin_part_gen}-\ref{eq:langevin_field_gen}) that the coupling strength is set by $hK\sim \rho_0^{1/2} V$ instead of $h$ itself. 
This coupling should then be compared to the thermal energy, so that the perturbative computation is valid when
\begin{equation}\label{eq:valid_pert}
\frac{\rho_0^{1/2}V}{T}\ll 1.
\end{equation}
For the perturbative computations to be valid together with the LDT (see Eq.~(\ref{eq:valid_ldt})), the pair interaction should be small and bounded, $V/T\ll 1$, meaning that the particles are \emph{soft}: they can overlap completely at a finite energy cost. We restrict ourselves to this case from now on.
An example of soft particles is given by polymer coils, whose effective pair potential is almost Gaussian~\cite{Louis2000b}.


Note that we consider here that the tracer is equivalent to the other particles, but it can also be different, representing for instance a hard sphere driven through soft polymer coils~\cite{Kruger2009}. This difference is easily integrated in the general formalism: if $U(\xx)$ is the interaction potential between the tracer and the bath particles, Eq. (\ref{eq:map_form}) is replaced by
\begin{equation}
\tilde K(\kk)=-\rho_0\tilde U(\kk).
\end{equation}

\subsection{Effective tracer evolution equation}\label{}

We show that an effective, non-Markovian, evolution equation can be written for the tracer. 
The field evolution equation (\ref{eq:langevin_field_gen}) is linear and can be integrated:
\begin{equation}\label{eq:solfield_gen}
\phi(\xx,t)=\int_{-\infty}^t \left(e^{-|t-t'|R\Delta}[hRK\delta_{\xx_0(t')}+\xi(\cdot,t')]\right)(\xx)\dd t'.
\end{equation}
The lower integration bound $t_0=-\infty$ signifies that the system has forgotten its initial configuration and can be considered in a stationary state.
Inserting this solution into the particle dynamics (\ref{eq:langevin_part_gen}), we get
\begin{equation}\label{eq:eff_part_dyn}
\dot\xx_0(t) =\ff\\ + \int_{-\infty}^t \FF(\xx_0(t)-\xx_0(t'),t-t')\dd t' + \eeta(t) + \XXi(\xx_0(t),t),
\end{equation}
where $\XXi(\xx,t)$ is a Gaussian noise with correlation function
\begin{equation}\label{eq:eff_part_noise}
\left\langle \XXi(\xx,t)\XXi(\xx',t')\transp \right\rangle=2T\GG(\xx-\xx',t-t').
\end{equation}
We have introduced the functions
\begin{eqnarray}
\FF(\xx,t) & = h^2\nnabla Ke^{-tR\Delta}RK(\xx),\\
\GG(\xx,t) & = -h^2\nnabla\nnabla\transp K^2 e^{-|t|R\Delta}\Delta^{-1}(\xx),
\end{eqnarray}
that read in Fourier space
\begin{eqnarray}
\tilde\FF(\kk,t) & = ih^2\kk \tilde K(\kk)^2 \tilde R(\kk) e^{-\tilde R(\kk)\tilde\Delta(\kk)t},\\
\tilde\GG(\kk,t) & = h^2\kk\kk\transp \tilde K(\kk)^2\tilde\Delta^{-1}(\kk) e^{-\tilde R(\kk)\tilde\Delta(\kk)|t|}.
\end{eqnarray}
For a tracer in a colloidal bath, they are
\begin{eqnarray}
\tilde\FF(\kk,t) & = i\rho_0\kk\kk^2 \tilde V(\kk)^2 e^{-\kk^2[T+\rho_0\tilde V(\kk)]t}, \label{eq:F_bath}\\
\tilde\GG(\kk,t) & = \rho_0\kk\kk\transp \frac{\tilde V(\kk)^2}{T+\rho_0\tilde V(\kk)} e^{-\kk^2[T+\rho_0\tilde V(\kk)]|t|}. \label{eq:G_bath}
\end{eqnarray}

The equations (\ref{eq:eff_part_dyn}, \ref{eq:eff_part_noise}, \ref{eq:F_bath}, \ref{eq:G_bath}) contain all the information on the tracer dynamics in a colloidal bath: they provide the explicit generalized Langevin equation of a tracer, biased or not, in a dense bath of soft colloidal particles.
This constitutes an important result of this article.
The second term on the right hand side of (\ref{eq:eff_part_dyn}) is a memory term; it represents the action of the tracer on its surrounding particles that propagates via the other particles and finally acts back on the tracer. The last term is a colored noise in time and space, coming from the white thermal noise on each particle that propagates through the bath before acting on the tracer.

\section{Application to a colloidal bath at equilibrium}\label{sec:tracer_no_force}

Before considering a tracer submitted to an external force in a colloidal bath, we show that the linearized Dean equation allows one to recover previous results in the physics of colloidal systems.

\subsection{Pair correlation function}\label{}

First, we compute the pair correlation function of the bath with the linearized Dean equation without taking the tracer apart (\ref{eq:dean_lin}); it is defined as~\cite{Hansen2006}
\begin{equation}
h(\xx)= \frac{\langle \phi(0,t)\phi(\xx,t) \rangle-\delta(\xx)}{\rho_0}.
\end{equation}
To compute the two-point correlation function of the field $\phi(\xx,t)$, we write the solution (\ref{eq:solfield_gen}) without the effect of the tracer and in Fourier space:
\begin{equation}
\tilde\phi(\kk,t)=\int_{-\infty}^t e^{-\tilde R(\kk)\tilde\Delta(\kk)|t-t'|}\tilde\xi(\kk,t')\dd t'.
\end{equation}
Using the noise correlation function (\ref{eq:field_noise}) and integrating over $t'$, we can get the equal time two-point correlation function:
\begin{equation}
\left\langle \tilde\phi(\kk,t)\tilde\phi(\kk',t) \right\rangle=\frac{(2\pi)^d T\delta(\kk+\kk')}{\tilde\Delta(\kk)}.
\end{equation}
The operator $\tilde\Delta(\kk)$ being given by the mapping (\ref{eq:map_delta}) we obtain the pair correlation function in Fourier space,
\begin{equation}\label{eq:deanlin_h}
\tilde h(\kk)=-\frac{T^{-1}\tilde V(\kk)}{1+\rho_0 T^{-1}\tilde V(\kk)};
\end{equation}
This formula is exactly the one obtained for the pair correlation function in the mean-field approximation~\cite{Lang2000,Likos2001}, also called \emph{random phase approximation}~\cite{Louis2000}. That shows that the linearized Dean equation contains the mean-field approximation.

We can also compute the average density in the reference frame of the tracer (\ref{eq:av_dens_tracer}) without the external force~$\ff$, $\langle \rho^*(\xx) \rangle=\rho_0+\rho_0^{1/2}\psi(\xx)$ with
\begin{equation}
\psi(\xx)=\left\langle  \phi(\xx+\xx_0(0), 0)\right\rangle.
\end{equation}
It is not possible to solve exactly the particle and field dynamics (\ref{eq:langevin_part_gen}, \ref{eq:langevin_field_gen}), we thus restrict ourselves to a perturbative computation in the coupling constant $h$.
To determine the profile $\psi(\xx)$ to the order $h$, we do not have to take into account the effect of the field on the particle motion, that is thus a simple Brownian motion.
We average the general solution for the field $\phi$ (\ref{eq:solfield_gen}); the average over the field noise gives $0$, so that the profile reads in Fourier space
\begin{equation}\label{eq:psi1_f0}
\tilde\psi(\kk)=h\int_0^\infty e^{-t\tilde R(\kk)\tilde\Delta(\kk)} \tilde R(\kk)\tilde K(\kk) \left\langle e^{-i\kk\cdot\xx_0(-t)} \right\rangle_0\dd t.
\end{equation}
The index $0$ on the average means that it is the average for the pure Brownian motion, without coupling between the particle and the field $\phi$. At a given time $t$, the position of the particle $\xx_0(-t)$ is a Gaussian random variable of zero mean and variance $2D_\xx t\un$, giving for the average
\begin{equation}\label{eq:average_exp_pos}
\left\langle e^{-i\kk\cdot\xx_0(-t)} \right\rangle_0=e^{-D_\xx \kk^2 t};
\end{equation}
an integration over $t$ gives the profile,
\begin{equation}\label{eq:av_prof_f0_gen}
\tilde \psi(\kk)=\frac{h\tilde R(\kk)\tilde K(\kk)}{\tilde R(\kk)\tilde \Delta(\kk) + D_\xx\kk^2}.
\end{equation}
In the case of a tracer in a bath of soft particles, it is,
\begin{equation}
\tilde \psi(\kk)=-\frac{\rho_0^{1/2}}{2T}\frac{\tilde V(\kk)}{1+\frac{\rho_0\tilde V(\kk)}{2T}}.
\end{equation}
leading to a density correction $\langle\delta\rho^*(\xx)\rangle=\langle \rho^*(\xx) \rangle-\rho_0$ given by
\begin{equation}\label{eq:dens_around_tracer_noforce}
\frac{\left\langle \widetilde{\delta\rho}^*(\kk) \right\rangle}{\rho_0}=-\frac{(2T)^{-1}\tilde V(\kk)}{1+\rho_0(2T)^{-1}\tilde V(\kk)}.
\end{equation}
We recover the pair correlation function (\ref{eq:deanlin_h}), the only difference being the factor 2. Due to the fluctuation-dissipation theorem~\cite{Hansen2006}, the pair correlation function $h(\xx)$ gives the density response at $\xx$ to the inclusion of a fixed particle at the origin. 
The density in the reference frame of the tracer is different, because the tracer diffuses. If the tracer is kept fixed when the profile is computed, the factor 2 is removed and the pair correlation function (\ref{eq:deanlin_h}) is recovered. 

The two results (\ref{eq:deanlin_h}) and (\ref{eq:dens_around_tracer_noforce}) ensure the relevance of the LDT~(\ref{eq:ldt1}-\ref{eq:ldt4}) to describe a dense liquid.

\subsection{Diffusion coefficient of a free tracer}\label{}

The effective diffusion coefficient of a free tracer (i.e. without forcing) was computed in the general case (\ref{eq:langevin_part_gen}-\ref{eq:field_noise}) in the limit of small coupling $h$ between the field and the tracer~\cite{Demery2011}. 
A one loop path integral computation gives the effective diffusion coefficient,
\begin{equation}
\frac{D\ind{eff}}{D_\xx}=1-\frac{h^2}{d}\int
\frac{\kk^2|\tilde K(\kk)|^2 }{\tilde\Delta(\kk) \left[\tilde R(\kk)\tilde\Delta(\kk) + T\kk^2\right]}\frac{\dd\kk}{(2\pi)^d}.
\end{equation}
For a tracer in a colloidal bath, it reads
\begin{equation}
\frac{D\ind{eff}}{D_\xx}=1\\- \frac{\rho_0}{2dT^2}\int \frac{\tilde V(\kk)^2}{\left[1+\rho_0T^{-1}\tilde V(\kk)\right]\left[1+\rho_0(2T)^{-1}\tilde V(\kk)\right]} \frac{\dd\kk}{(2\pi)^d}.
\end{equation}
Interestingly, we recover the result of Dean and Lefèvre~\cite{Dean2004} (see Eq.~(47)) for the same system but derived in a very different manner. 
The method used in~\cite{Dean2004} avoids the use of the Dean equation and deals directly with the perturbative computation of the probability density function of the $N+1$ particles, that obeys a Fokker-Planck equation in a $(N+1)\times d$-dimensional space.
This technique is valid for soft particles, but does not require a high density, contrarily to ours, suggesting that the LDT may be valid for moderate densities.

The compatibility of the two approaches confirms the pertinence of the LDT to describe the motion of a tracer in a colloidal bath.

\section{Application to a tracer submitted to a constant force}\label{sec_mob_diff}

We now study the out of equilibrium configuration where a constant force $\ff$ is applied to the tracer and compute the stationary density profile around the tracer, the tracer effective mobility and its diffusion coefficient, perturbatively in the coupling $h=\rho_0^{-1/2}$, still assumed to be small.



\subsection{Stationary field profile}\label{}

It is interesting to see how the tracer affects the surrounding particle density when it moves. We only need to generalize the profile (\ref{eq:av_prof_f0_gen}) obtained without applied force. 
Now, the bare particle motion is a biased Brownian motion and the particle position $\xx_0(t)$ is a random Gaussian variable with mean $\ff t$ and variance $2D_\xx t\un$. The analog of~(\ref{eq:average_exp_pos}) is now
\begin{equation}
\left\langle e^{-i\kk\cdot\xx_0(-t)} \right\rangle_0=e^{-\left(D_\xx \kk^2-i\ff\cdot\kk\right) t},
\end{equation}
leading to the profile in Fourier space
\begin{equation}
\tilde \psi(\kk)=\frac{h\tilde R(\kk)\tilde K(\kk)}{\tilde R(\kk)\tilde \Delta(\kk) + D_\xx\kk^2 -i\ff\cdot\kk}.
\end{equation}

\begin{figure}
\begin{center}
\includegraphics[width=0.6\linewidth]{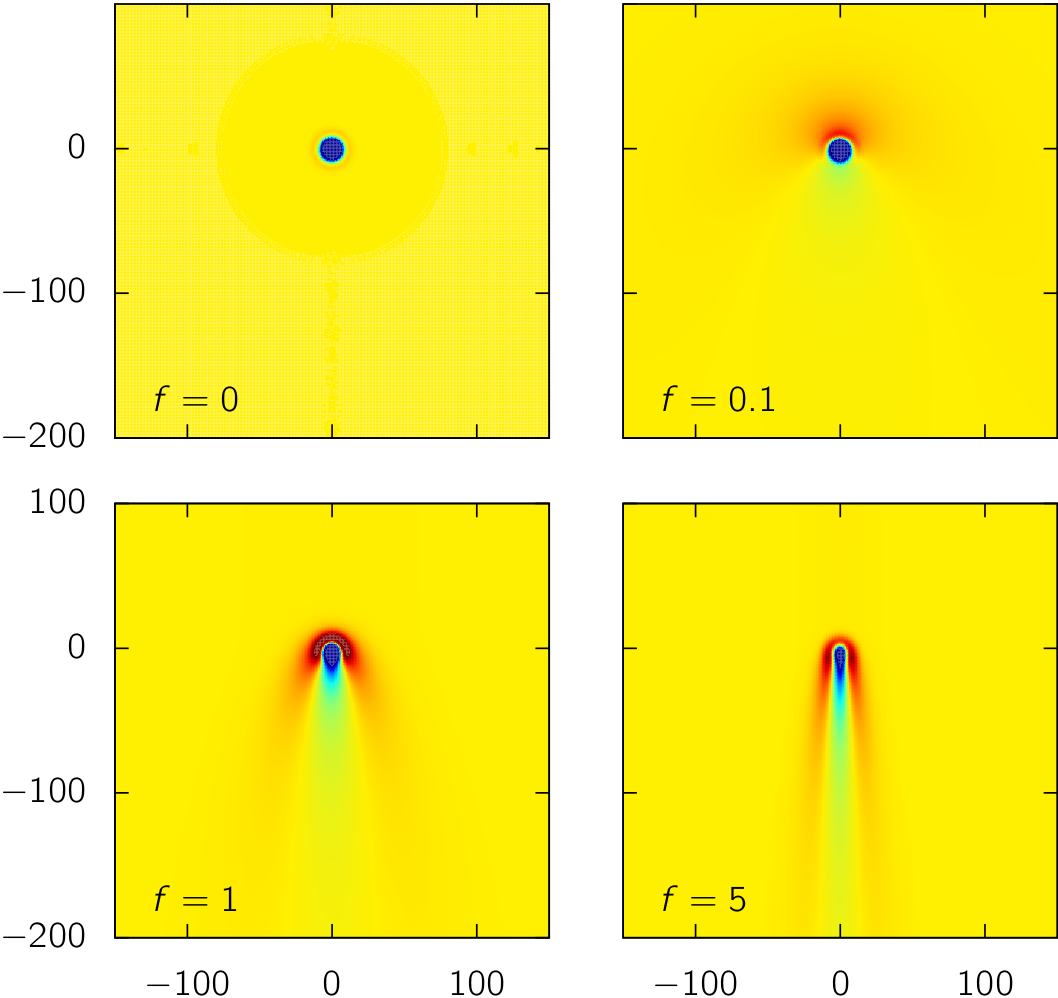}
\end{center}
\caption{(Colour online) Stationary bath density in the reference frame of the tracer for different values of the applied force.}
\label{fig_profils2d}
\end{figure}

In the case of a tracer in a bath of soft particles, it is
\begin{equation}\label{eq:profile_force}
\tilde \psi(\kk)=-\frac{\rho_0^{1/2}}{2T}\frac{\kk^2\tilde V(\kk)}{\left[1+\frac{\rho_0\tilde V(\kk)}{2T} \right]\kk^2 -i\frac{ \ff\cdot\kk}{2T}}.
\end{equation}
The full real space profile cannot be obtained analytically, but the Fourier transform can be performed numerically. Examples for Gaussian spheres ($V(\xx)= \exp \left(-\xx^2/2\right)$) in dimension $d=2$ are given in Figure~\ref{fig_profils2d}.

\begin{figure}
\begin{center}
\includegraphics[width=0.6\linewidth]{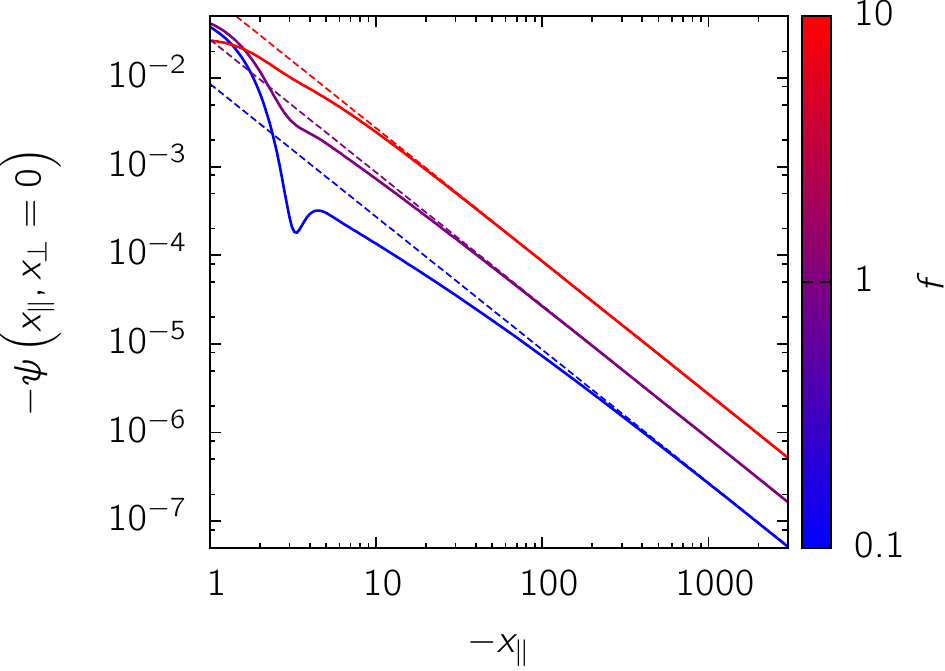}
\end{center}
\caption{(Colour online) Stationary profile behind the particle in dimension $d=2$ along the direction of the force, for driving forces $f=0.1,\, 1,\, 10$. Solid lines are numerical computations and dashed lines the asymptotic analytical prediction (\ref{eq:profil_asympt}).}
\label{fig_profils_x}
\end{figure}

In dimension $d\geq 2$ and for non zero bias, the profile is singular in Fourier space at $\kk=\zero$: the limits $k_\parallel\to 0$ and $\kk_\perp\to\zero$ do not commute (these directions are defined with respect to the force $\ff$).
This may lead to an algebraic decay of the profile in real space. The algebraic terms in real space can be computed by keeping only the singular term in Fourier space, that is a function $\tilde\psi\ind{sing}(\kk)$ such that $\tilde\psi(\kk)-\tilde\psi\ind{sing}(\kk)$ is regular at $\kk=\zero$. The singular part is not unique, but it is defined up to an additive regular part that decays faster than algebraicaly in real space. This freedom in the singular part allows us to pick a simple one that can be Fourier transformed exactly:
\begin{equation} 
\tilde\psi\ind{sing}(\kk)=-\frac{i\rho_0^{1/2}\tilde V(0)}{2T+\rho_0\tilde V(0)}\frac{fk_\parallel}{[2T+\rho_0\tilde V(0)]\kk_\perp^2-ifk_\parallel}.
\end{equation}
To Fourier transform this expression, the integration over $k_\parallel$ can be performed using the Residue theorem and it remains a second derivative of a Gaussian. 
We find that if $x_\parallel>0$, i.e. in front of the tracer, $\psi\ind{sing}(\xx)=0$. On the other hand, behind the tracer where $x_\parallel<0$, we obtain an algebraic decay as
\begin{equation}\label{eq:profil_asympt}
\frac{ \left\langle\delta\rho^*\ind{sing}(x_\parallel,\xx_\perp=\zero)  \right\rangle}{\rho_0}\underset{x_\parallel\rightarrow -\infty}{\sim}
-\frac{(d-1)\tilde V(0) f^\frac{d-1}{2}}{2^d\pi^\frac{d-1}{2}\left[2T+\rho_0\tilde V(0) \right]^\frac{d+1}{2}}
\times\frac{1}{|x_\parallel|^\frac{d+1}{2}}.
\end{equation}
This expression compares nicely to numerical inversion of the Fourier transform of the complete expression (\ref{eq:profile_force}) for $d=2$ as is shown on Figure~\ref{fig_profils_x}.
Unexpectedly, the same algebraic decay was observed for a tracer driven through a hard core lattice gas in dimension $d=2$~\cite{Benichou2000}. We see later that the correspondence between the two systems is even deeper.

In dimension $d=1$, the profile (\ref{eq:profile_force}) is not singular in Fourier space: the real space profile decays faster than algebraically.
This comes from the fact that 
we deal here with soft particles, that can cross at a finite energy cost.

\subsection{Path-integral representation}\label{}

The path-integral representation introduced in~\cite{Demery2011} may be extended to compute the correction to the bare mobility and diffusion coefficient to the order $h^2$ when an external force is applied to the tracer.
The effective dynamics~(\ref{eq:eff_part_dyn}) can be mapped to a field theory~\cite{Demery2011, Aron2010} with action
\begin{equation}
S[\xx,\pp]=S_0[\xx,\pp]+S\ind{int}[\xx,\pp],
\end{equation}
where $S_0[\xx,\pp]$ is the action of the bare particle,
\begin{equation}
S_0[\xx,\pp]=-i\int \pp(t)\cdot \left[\dot\xx(t)-\ff \right]\dd t+ D_\xx\int|\pp(t)|^2 \dd t,
\end{equation}
and $S\ind{int}[\xx,\pp]$ is the action carrying the particle-field interaction,
\begin{eqnarray}
\fl S\ind{int}[\xx,\pp]=i\int \pp(t)\cdot\FF(\xx(t)-\xx(t'),t-t')\theta(t-t')\dd t\dd t' \nonumber \\
 + T \int \pp(t)\cdot\GG(\xx(t)-\xx(t'),t-t')\pp(t')\theta(t-t')\dd t\dd t'.
\end{eqnarray}
We introduced the response field $\pp(t)$, $\theta(t)$ is the Heaviside function and the index 0 for the tracer position is dropped from now on.
We use the It\=o convention~\cite{Aron2010}.

The idea is to treat the interaction action, that is proportional to $h^2$, perturbatively: 
we write for an observable $O[\xx]$
\begin{equation}
\left\langle O[\xx] \right\rangle =\frac{\left\langle O[\xx]\exp \left(-S\ind{int}[\xx,\pp] \right) \right\rangle_0}{\left\langle \exp \left(-S\ind{int}[\xx,\pp] \right) \right\rangle_0}
\simeq \frac{\left\langle O[\xx]\left(1-S\ind{int}[\xx,\pp] \right) \right\rangle_0}{\left\langle 1-S\ind{int}[\xx,\pp] \right\rangle_0}. \label{eq:moy_obs}
\end{equation}
The index $0$ in averages indicates that they are computed with the bare action $S_0[\xx,\pp]$. Since the bare action is quadratic, only the two first moments are needed to compute all the averages with the Wick's theorem~\cite{Wick1950}. They can be computed using the Schwinger-Dyson equation~\cite{Dyson1949b}, as in~\cite{Demery2011}, and we get
\begin{eqnarray}
\langle \xx(t) \rangle_0 & =\ff t, \label{eq:av_pos}\\
\langle \pp(t) \rangle_0 & = \zero,\\
\left\langle \pp(t)\pp(t')\transp \right\rangle_0 & = \zero,\\
\left\langle \xx(t)\pp(t')\transp \right\rangle_0 & = i \chi_{[0,t)}(t'),\\
\left\langle [\xx(t)-\ff t][\xx(t')-\ff t']\transp \right\rangle_0 & = 2T \mathrm{L}([0,t)\cap[0,t')).
\end{eqnarray}
We use $\chi_A(t)$ as the characteristic function of the interval $A$ and $\mathrm{L}(A)$ as its length. If $O_j$ are linear observables in $\xx$ and $\pp$, the Wick's theorem allows to show that
\begin{equation}
\fl\left\langle \prod_{j=1}^n O_j e^{i\kk\cdot\xx} \right\rangle_0=e^{i\kk\cdot \langle \xx \rangle-\frac{1}{2}\kk\transp \left\langle \xx\xx\transp \right\rangle_0\kk} \sum_{J\subset N} \left(\prod_{j\in J} (i\kk)\cdot \left\langle O_j\xx \right\rangle_0 \left\langle \prod_{j\notin J} O_j \right\rangle_0\right),
\end{equation}
where the sum over $J$ is the sum over all subsets of $N=\{1,\dots,n\}$.

We start with $\left\langle S\ind{int}[\xx,\pp] \right\rangle_0$, that contains
\begin{equation}
\left\langle \pp(t) e^{i\kk\cdot[\xx(t)-\xx(t')]} \right\rangle_0=\zero
\end{equation}
and
\begin{equation}
\left\langle \pp(t)\pp(t')\transp e^{i\kk\cdot[\xx(t)-\xx(t')]} \right\rangle_0=\zero,
\end{equation}
so that
\begin{equation}
\left\langle S\ind{int}[\xx,\pp] \right\rangle_0=0.	
\end{equation}
Together with (\ref{eq:moy_obs}), this means that the average of an observable $O[\xx]$ is, to the second order in $h$, 
\begin{equation}\label{eq:moy_obs2}
\langle O[\xx] \rangle\simeq \langle O[\xx] \rangle_0- \langle O[\xx]S\ind{int}[\xx,\pp] \rangle_0.
\end{equation}

To compute the mobility, we need $\left\langle \xx(t) S\ind{int}[\xx,\pp] \right\rangle_0$; it invokes, for $t'>t''$,
\begin{equation}
\left\langle \xx(t) \pp(t')\transp e^{i\kk\cdot[\xx(t')-\xx(t'')]} \right\rangle_0\\=i\chi_{[0,t)}(t') e^{i\kk\cdot\ff (t'-t'')-T\kk^2|t'-t''|},
\end{equation}
and
\begin{equation}
\fl\left\langle \xx(t) \pp(t')\transp\pp(t'') e^{i\kk\cdot[\xx(t')-\xx(t'')]} \right\rangle_0=-i\chi_{[0,t)}(t') \kk e^{i\kk\cdot\ff (t'-t'')-T\kk^2|t'-t''|}.
\end{equation}
Using the expression of the interaction action and integrating over the interaction times $t'$ and $t''$ gives the long time behavior
\begin{equation} \label{eq:dev_av_pos}
\fl\left\langle \xx(t) S\ind{int}[\xx,\pp] \right\rangle_0 \underset{t\rightarrow\infty}{\sim} 
t\ff h^2 \int \frac{k_\parallel^2|\tilde K(\kk)|^2 \left[\tilde R(\kk)\tilde \Delta(\kk)+T\kk^2 \right]}{\tilde\Delta(\kk)\left(\left[\tilde R(\kk)\tilde\Delta(\kk)+T\kk^2\right]^2+[\ff\cdot\kk]^2\right)} \frac{\dd\kk}{(2\pi)^d}.
\end{equation}

To compute the effective diffusion coefficient, we need $\left\langle [\xx(t)-\ff t]^2 S\ind{int}[\xx,\pp] \right\rangle_0$ and notably
\begin{eqnarray}
\fl\left\langle [\xx(t)-\ff t]^2 \pp(t')\transp e^{i\kk\cdot[\xx(t')-\xx(t'')]} \right\rangle_0 \nonumber \\
= -4T\kk \transp \chi_{[0,t)}(t')(t'-t'')e^{i\kk\cdot\ff (t'-t'')-T\kk^2|t'-t''|},
\end{eqnarray}
and
\begin{eqnarray}
\fl\left\langle [\xx(t)-\ff t]^2 \pp(t')\transp\pp(t'') e^{i\kk\cdot[\xx(t')-\xx(t'')]} \right\rangle_0 \nonumber\\
= 2\chi_{[0,t)}(t') \left[2T\kk^2(t'-t'')-\chi_{[0,t)}(t'') \right]  e^{i\kk\cdot\ff (t'-t'')-T\kk^2|t'-t''|}.
\end{eqnarray}
In the long time limit, we get
\begin{eqnarray}
\fl \left\langle [\xx(t)-\ff t]^2 S\ind{int}[\xx,\pp] \right\rangle_0=2t h^2T \nonumber\\
 \fl \quad\times \int
\frac{\kk^2|\tilde K(\kk)|^2 \left[\tilde R(\kk)\tilde\Delta(\kk) + T\kk^2\right] \left[\left(\tilde R(\kk)\tilde\Delta(\kk) + T\kk^2\right)^2-3(\ff\cdot\kk)^2 \right]}{\tilde\Delta(\kk) \left[\left(\tilde R(\kk)\tilde\Delta(\kk) + T\kk^2\right)^2 + (\ff\cdot\kk)^2\right]^2}\frac{\dd\kk}{(2\pi)^d}.
\end{eqnarray}
These averages now allow us to compute the perturbative corrections to the bare mobility and diffusion coefficient.

\subsection{Mobility}\label{}

The average position is obtained by inserting the averages (\ref{eq:av_pos}) and (\ref{eq:dev_av_pos}) in the perturbative expansion (\ref{eq:moy_obs2}); it finally leads the mobility, defined in (\ref{eq:def_mobility}):
\begin{equation}\label{eq:keff}
\kappa\ind{eff} = 1 - h^2 \int \frac{k_\parallel^2|\tilde K(\kk)|^2[\tilde R(\kk)\tilde\Delta(\kk)+T\kk^2]}{\tilde\Delta(\kk) \left[\left(\tilde R(\kk)\tilde\Delta(\kk) + T\kk^2\right)^2 + (\ff\cdot\kk)^2\right]}\frac{\dd\kk}{(2\pi)^d}.
\end{equation}
The correction to the bare mobility is negative and its absolute value decreases as the applied force increases.
The physical interpretation is that when the force it too large, the tracer moves rapidly and the field does not have any time to respond to the presence of the tracer.
Another consequence of this effect is the non-monotonic behavior of the drag force as a function of the imposed velocity~\cite{Demery2010, Demery2010a, Kadau2008}. 

In the case of a tracer in a colloidal bath, the effective mobility reads
\begin{equation}\label{eq:keff_spheres}
\fl\kappa\ind{eff}=1-\frac{\rho_0}{2T^2} \int \frac{k_\parallel^2\kk^2|\tilde V(\kk)|^2 \left(1+\frac{\rho_0\tilde V(\kk)}{2T} \right)}{\left(1+\frac{\rho_0\tilde V(\kk)}{T} \right)\left[\left(1+\frac{\rho_0\tilde V(\kk)}{2T} \right)^2\kk^4+\left( \frac{ \ff\cdot\kk}{2 T} \right)^2\right]}\frac{\dd\kk}{(2\pi)^d}.
\end{equation}
We note that the correction to the bare mobility $\kappa_\xx=1$ is a non monotonic function of the bath density $\rho_0$.

\subsection{Diffusion coefficient}\label{}

The effective diffusion coefficient is defined by
\begin{eqnarray}
\left\langle [\xx(t)- \langle \xx(t) \rangle]^2 \right\rangle & =\left\langle [\xx(t)-\ff t- \langle \xx(t)-\ff t \rangle]^2 \right\rangle \nonumber \\
& = \left\langle [\xx(t)-\ff t]^2 \right\rangle - \left\langle \xx(t)-\ff t \right\rangle^2.
\end{eqnarray}
From (\ref{eq:dev_av_pos}) the second term on the right hand side is of order $h^4$, and thus does not contribute to the order $h^2$. We thus need only the first term to compute the effective diffusion coefficient to the order $h^2$.

In the general case, the effective diffusion coefficient reads
\begin{eqnarray}\label{eq:deff_gen}
\fl \frac{D\ind{eff}}{D_\xx} = 1 \nonumber \\ 
\fl\quad - \frac{h^2}{d}\int
\frac{\kk^2|\tilde K(\kk)|^2 [\tilde R(\kk)\tilde\Delta(\kk) + T\kk^2]\left([\tilde R(\kk)\tilde\Delta(\kk) + T\kk^2]^2-3(\ff\cdot\kk)^2\right)}{\tilde\Delta(\kk) \left[(\tilde R(\kk)\tilde\Delta(\kk) + T\kk^2)^2 + (\ff\cdot\kk)^2\right]^2}\frac{\dd\kk}{(2\pi)^d}.
\end{eqnarray}
First, we note that without forcing ($\ff=\zero$), the system is at equilibrium and the Einstein relation is satisfied:
\begin{equation}
D\ind{eff}=T\kappa\ind{eff}.
\end{equation}
Second, when the force is high enough, i.e. when the system is far from equilibrium, the correction to the bare diffusion coefficient $D_\xx$ is positive. This is a pure out of equilibrium effect, since it has been shown in~\cite{Dean2011} that the diffusion coefficient is always reduced by its interaction with the surrounding field when the system is at equilibrium. 
It also recalls the effect seen in~\cite{Demery2011}, where the system was driven out of equilibrium by breaking the detailed balance in its dynamics; in this case, it was also possible to observe enhanced diffusion. 

The effective diffusion coefficient of a tracer in a colloidal bath is
\begin{equation}\label{eq:deff_spheres}
\fl \frac{D\ind{eff}}{D_\xx}=1-\frac{\rho_0}{2dT^2}\int
\frac{\kk^4|\tilde V(\kk)|^2 \left[1+\frac{\rho_0\tilde V(\kk)}{2T} \right]\left(\left[1+\frac{\rho_0\tilde V(\kk)}{2T} \right]^2\kk^4 - 3\left[ \frac{ \ff\cdot\kk}{2T} \right]^2\right)}{\left[1+\frac{\rho_0\tilde V(\kk)}{T} \right] \left(\left[1+\frac{\rho_0\tilde V(\kk)}{2T} \right]^2\kk^4 + \left[\frac{ \ff\cdot\kk}{2 T} \right]^2\right)^2}\frac{\dd\kk}{(2\pi)^d}.
\end{equation}

The bias enhanced diffusion was also observed for a tracer in a dense hard core lattice gas~\cite{Benichou2013b, Benichou2013c, Benichou2013e}. 
Similarly to the mobility, the density dependence is non trivial: the correction is linear in the density at small density, and decays as $\delta D\sim\rho_0^{-1}$ at high density. Strikingly, the correction can turn from positive to negative when the density increases, an effect seen in a hard core lattice gas~\cite{Benichou2013d}.

The correspondence between a hard core lattice gas, where the particles are very hard and their motion is severely constrained, and the soft colloids studied here, was not expected.
The fact that the same effects can be observed in the two opposite limits proves their robustness. 

\section{Numerical simulations}\label{sec:numerics}

To test our predictions, we have performed one dimensional molecular dynamics simulations of Eq.~(\ref{eq:langevin_npart}) for harmonic spheres, corresponding to the pair potential
\begin{equation}
V(\xx)=(1-|\xx|)^2\theta(1-|\xx|).
\end{equation}
Since this potential is bounded, the particles can cross and the dimension $d=1$ is not singular. 
This is very different from hard core particles, that are singular in one dimension; in this case, the tracer would undergo single-file diffusion, characterized by a sub-diffusive behavior~\cite{Wei2000}.

\begin{figure}
\begin{center}
\includegraphics[width=0.6\linewidth]{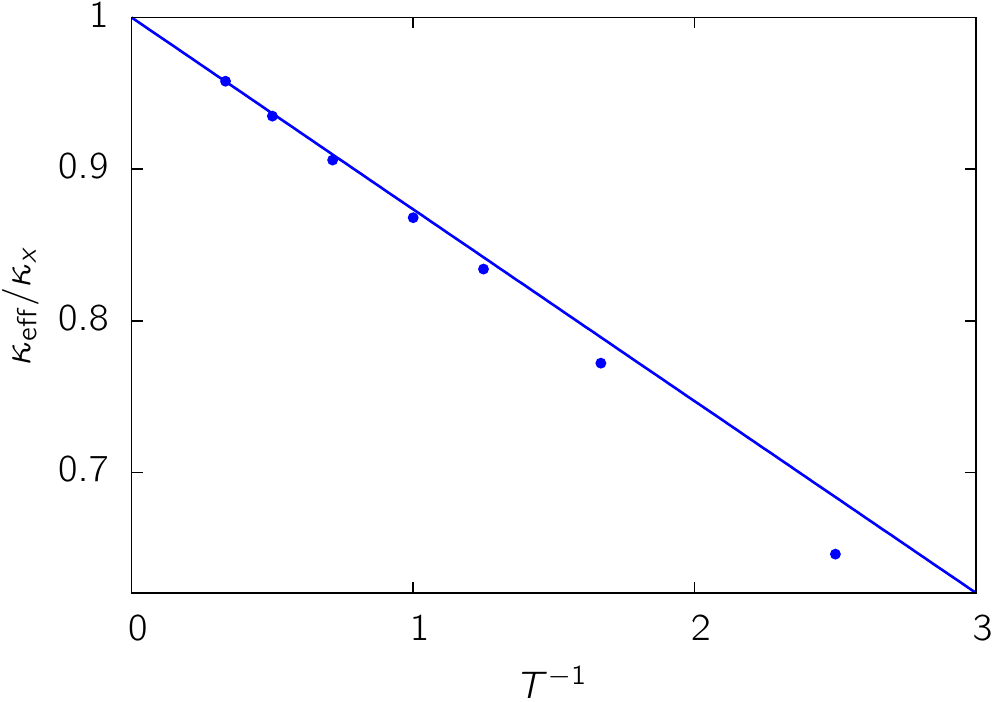}
\end{center}
\caption{(Colour online) Effective mobility as a function of the inverse temperature for $\rho_0/T=5$ and $f/T=1$. The points are the simulations results and the line is the analytical prediction~(\ref{eq:keff_spheres}).}
\label{fig:rho_keff}
\end{figure}

The ratio of the effective dynamical quantities (mobility and diffusion coefficient) over the bare ones depend on three variables: the density $\rho_0$, the temperature $T$ and the force $f$ applied on the tracer. More precisely, the prediction for the effective mobility (\ref{eq:keff_spheres}) takes the form
\begin{equation}
\frac{\kappa\ind{eff}}{\kappa_\xx}=1-T^{-1}g_\kappa \left(\frac{\rho_0}{T},\frac{f}{T} \right).
\end{equation}
The same relation holds for the diffusion coefficient, with a different scaling function $g_D$. 
The evolution of the effective mobility with the temperature for $\rho_0/T=5$ and $f/T=1$ is plotted on Figure~\ref{fig:rho_keff}. The agreement between the theory and the simulations is excellent up to a correction of $30\,\%$ with respect to the bare value. 

\begin{figure}
\begin{center}
\includegraphics[width=0.6\linewidth]{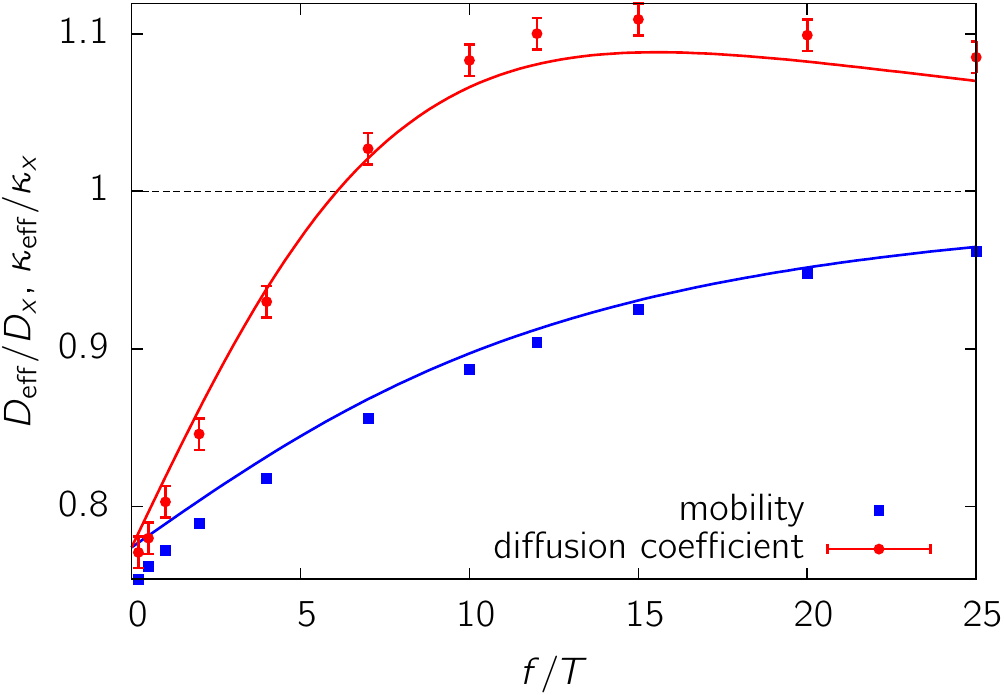}
\end{center}
\caption{(Colour online) Effective mobility and diffusion coefficient as a function of the applied force for $T=0.6$ and $\rho_0=3$. Lines are the analytical predictions (\ref{eq:keff_spheres}) and (\ref{eq:deff_spheres}).}
\label{fig:f_keff_deff}
\end{figure}

The dependence on the external force is shown on Figure~\ref{fig:f_keff_deff}. 
The agreement between our perturbative computations and the simulations is very good except for the diffusion coefficient at large forces $f/T\gtrsim 5$, where our computations underestimate the diffusion coefficient.
However, the diffusion enhancement at large forces is very satisfactorily captured by our analytical computation.

\begin{figure}
\begin{center}
\includegraphics[width=0.6\linewidth]{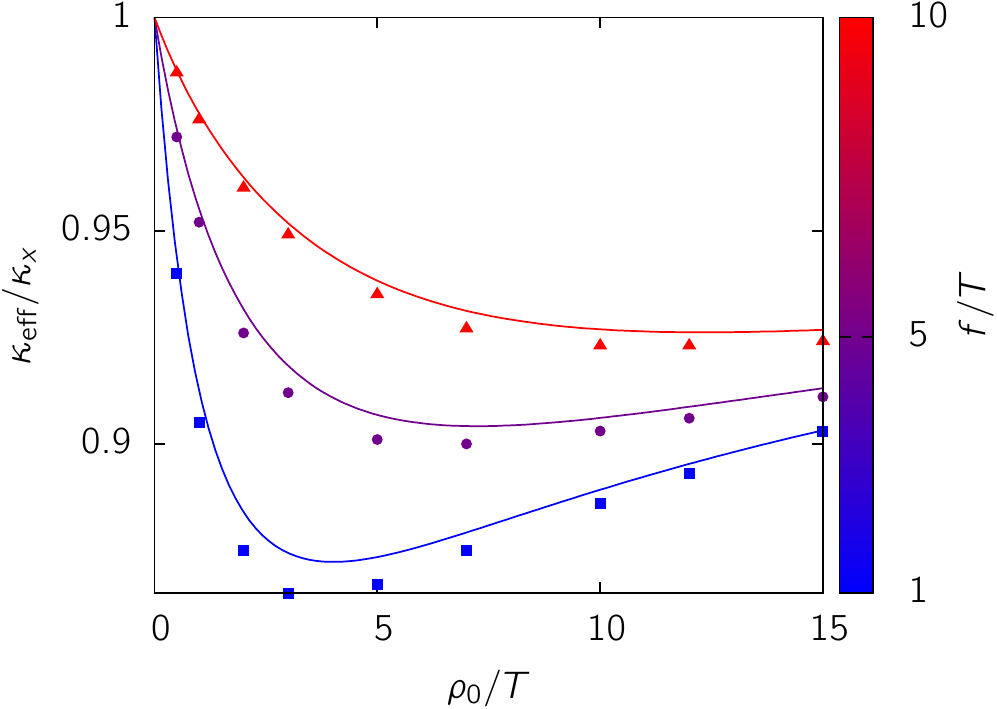}
\end{center}
\caption{(Colour online) Effective mobility as a function of the density $\rho_0/T$ for $T=1$ and different values of the force $f/T=1,5,10$.}
\label{fig:rho_Tconst_keff}
\end{figure}

\begin{figure}
\begin{center}
\includegraphics[width=0.6\linewidth]{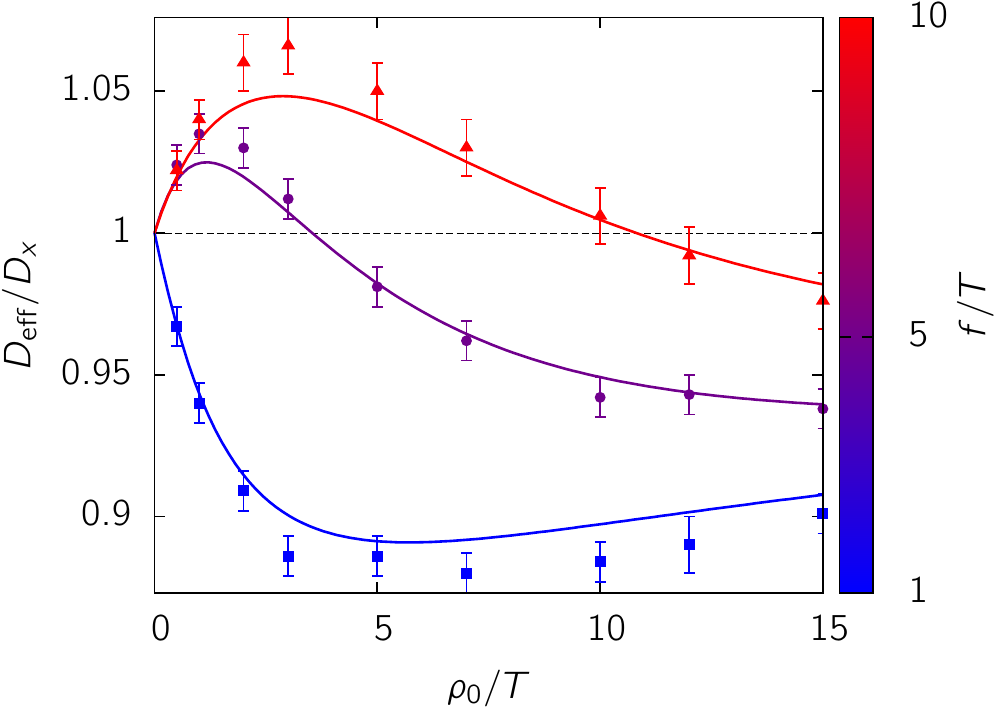}
\end{center}
\caption{(Colour online) Effective diffusion coefficient as a function of the density $\rho_0/T$ for $T=1$ and different values of the force $f/T=1,5,10$.}
\label{fig:rho_Tconst_deff}
\end{figure}

Last, we address the effect of the density $\rho_0$ at constant temperature and for different values of the force $f/T$, on Figure~\ref{fig:rho_Tconst_keff} for the mobility and on Figure~\ref{fig:rho_Tconst_deff} for the diffusion coefficient. 
The main effect is captured by our analytical results: at small forces, the diffusion coefficient is always reduced whereas at large forces, the diffusion coefficient increases at low densities. Increasing the density leads to a negative correction to the diffusion coefficient. 
A significant discrepancy is found at low densities for the diffusion coefficient; this is not surprising since the LDT validity condition (\ref{eq:valid_ldt}) is violated in this parameter regime.
The curves for the diffusion coefficient resemble strongly those obtained for a hard core lattice gas~\cite{Benichou2013d} (see Figure~2), confirming the universality of this effect.

We have obtained a good agreement for densities $\rho_0\simeq 4$ and $\rho_0^{1/2}V/T\simeq 2$. Comparing these values to the validity conditions of the LDT (\ref{eq:valid_ldt}) and the perturbative computation (\ref{eq:valid_pert}) leads us to the conclusion that these conditions are not too strict and that the results derived here are valid for a broad range of parameters.


\section{Conclusion}\label{sec_conclu}

We addressed the effective dynamical properties of a tracer in a bath of soft spheres. 
First, starting from the exact evolution equation of the overall density, namely the Dean equation, we took the tracer apart to obtain an equation describing the tracer motion in interaction with the overall density of the other particles.
If the bath is dense ($\rho_0\gg 1$), the evolution of the density fluctuations can be linearized.
This leads to a set of equations that we called the linearized Dean equation with a tracer~(LDT)~(\ref{eq:ldt1}-\ref{eq:ldt4}), that can be cast in a very general formalism describing the motion of a tracer interacting with a fluctuating field.
To prove the relevance of the LDT to describe dense liquids, we reproduced the pair correlation function of a fluid obtained by the mean-field treatment~\cite{Likos2001}. Moreover, using general results for the diffusion coefficient of a tracer interacting with a fluctuating field~\cite{Demery2011}, we recovered the tracer diffusion coefficient obtained by a completely different procedure~\cite{Dean2004}. 
In themselves, the LDT and the effective equation of motion for the tracer (\ref{eq:eff_part_dyn}, \ref{eq:eff_part_noise}, \ref{eq:F_bath}, \ref{eq:G_bath}) are very promising tools to study the single particle dynamics in a dense colloidal bath. 
Their simplicity makes them easy to adapt to various bath properties; it would for example be interesting to use them to model diffusion in the cytoplasm~\cite{Tabei2013}.


In a second part, we studied the effect of a constant force applied to the tracer, that drives the system out of equilibrium. We computed the average density profile around the tracer, and, using a path-integral formalism, the effective mobility and diffusion coefficient of the tracer. Our computations are perturbative in the particles pair interaction, and are thus valid for soft particles that can easily overlap (the validity condition is given by (\ref{eq:valid_pert})).
Our main predictions are: (i) The tracer diffusion is enhanced by the application of large external forces, (ii) The corrections to the mobility and the diffusion coefficient have a non-monotonic behavior and can even change sign as a function of the bath density. These predictions have been successfully tested against molecular dynamics simulations.
Surprisingly, they share many features with those derived in the opposite limit of hard core particles on a lattice: the tail of the density profile has the same exponent~\cite{Benichou2000} and the dependence of the mobility and the diffusion coefficient on the applied force and the bath density are very close~\cite{Benichou2013d}.
This qualitative agreement between results obtained for soft and hard particles suggests that the model proposed here can apply qualitatively to a wide range of parameters and proves the robustness of the observed effects.

Applying a constant force on the tracer is a way to drive the system out of equilibrium; another is to break the dynamics detailed balance, for example by reducing the effect of the tracer on its environment~\cite{Naji2009, Demery2011}. In both cases, the tracer diffusion coefficient may be enhanced by its environment. 
On the other hand, it has been shown that at equilibrium the tracer diffusion coefficient is always reduced by its interaction with the environment~\cite{Dean2011}. 
The augmentation of the diffusion coefficient under an external force is thus a pure and generic out of equilibrium effect.


The density dependence of the corrections to the dynamical quantities is noteworthy. As can be seen on Figure~\ref{fig:rho_Tconst_keff}, the correction to the bare mobility reaches a maximum for some value $(\rho_0/T)^*$ of the density-temperature ratio. 
This effect is likely to be the dynamical counterpart of the peak observed in the spatial structure when the density is varied that has been observed in the context of the jamming of soft spheres at low temperature~\cite{Jacquin2010, Jacquin2011a}.

Finally, we give a simple way to extend our model to be able to deal with an asymmetric configuration where the tracer is different from the other particles; that would for example be relevant to microrheology~\cite{Wilson2011b}. As noted above, it is enough to replace the operator $K(\xx)$ in (\ref{eq:langevin_part_gen}-\ref{eq:langevin_field_gen}) by $K(\xx)=-\rho_0 U(\xx)$, where $U(\xx)$ is the interaction potential between the tracer and the bath particles. 
The effect on the results turns out to be particularly simple: one just has to replace $|\tilde V(\kk)|^2$ by $|\tilde U(\kk)|^2$ in the integrand numerator of (\ref{eq:keff_spheres}) and (\ref{eq:deff_spheres}).

\section*{Acknowledgments}\label{}

The authors would like to thank David Dean and Vincent Krakoviack for illuminating discussions.
V. D. and O. B. acknowledge support from ERC Starting Grant No. FPTOpt-277998, and H. J. acknowledges funding from the ERC Grant OUTEFLUCOP.
\appendix

\section{Extraction of one particle from the Dean equation}\label{ap_dean_1p}

We show how to extract the tracer $i=0$ from the Dean equation and to write the evolution equation of the remaining particles density $\rho(\xx,t)=\sum_{j\geq 1}\delta(\xx-\xx_j(t))$. We start with Eq. (9) of~\cite{Dean1996} for the density associated to the particle~$i$, defined by $\rho_i(\xx,t))=\delta(\xx-\xx_i(t))$,
\begin{equation}\label{eq:dean_1pdens}
\partial_t\rho_i=T\nnabla^2\rho_i+\nnabla\cdot \left[\rho_i\nnabla \left(\sum_j V*\delta_{\xx_j} \right) \right]-\nnabla\cdot (\rho_i\eeta_i).
\end{equation}
The star $*$ denotes the convolution and the sum over $j$ runs over all the particles of the system; this sum can be written
\begin{equation}
\sum_j V*\delta_{\xx_j}=V*\rho\ind{tot},
\end{equation}
where $\rho\ind{tot}=\rho + \delta_{\xx_0}$ is the total density.
Now, summing (\ref{eq:dean_1pdens}) over all the particles but the tracer, we get
\begin{equation}
\partial_t\rho=T\nnabla^2\rho + \nnabla\cdot \left[\rho\nnabla(V*\rho\ind{tot}) \right] - \sum_{j\geq 1}\nnabla\cdot(\rho_j\eeta_j).
\end{equation}
It is then shown in~\cite{Dean1996} that the noise term can be rewritten
\begin{equation}
- \sum_{j\geq 1}\nnabla\cdot\left[\rho_j(\xx,t)\eeta_j(t)\right]=\nnabla\cdot \left[\rho(\xx,t)^{1/2}\xxi(\xx,t) \right]
\end{equation}
where $\eeta(\xx,t)$ is a Gaussian noise with correlation function
\begin{equation}
\left\langle \xxi(\xx,t)\xxi(\xx',t') \transp\right\rangle=2T\delta(\xx-\xx')\delta(t-t')\un.
\end{equation}
This leads to (\ref{eq:dean}).

\section{Functional operators in real and Fourier space}\label{ap:operators}

In this appendix, we define our notations and recall some basic properties of functional operators. We start in real space, and then see how it transposes to Fourier space. All the operators considered here are real. 

For two functions $f(\xx)$ and $g(\xx)$ and two operators $A(\xx,\xx')$ and $B(\xx,\xx')$, the scalar product of $f$ and $g$, the action of $A$ on $f$ and the product of $A$ and $B$ are respectively defined by
\begin{eqnarray}
f\cdot g & = \int f(\xx)g(\xx)\dd \xx,\label{eq:prod_scal}\\
(Af)(\xx) & = \int A(\xx,\xx')f(\xx')\dd \xx',\\
(AB)(\xx,\xx') & = \int A(\xx,\xx'')B(\xx'',\xx')\dd \xx''.\label{eq:prod_op}
\end{eqnarray}


An operator $A$ is invariant by translation if there exists a function $a$ such that
\begin{equation}
A(\xx,\xx')=a(\xx-\xx').
\end{equation}
Such an operator is isotropic if it only depends on the distance between $\xx$ and $\xx'$, $A(\xx,\xx')=a\left(|\xx-\xx'|\right)$.

We now switch to the Fourier space, with the Fourier transform defined by
\begin{eqnarray}
f(\xx) & = \int e^{i\kk\cdot\xx} \tilde f(\kk)\frac{\dd\kk}{(2\pi)^d},\\
A(\xx,\xx') & = \int e^{i(\kk\cdot\xx+\kk'\cdot\xx')} \tilde A(\kk,\kk')\frac{\dd\kk\dd\kk'}{(2\pi)^{2d}},
\end{eqnarray}
for a function and an operator, respectively. This definition allows us to translate (\ref{eq:prod_scal}-\ref{eq:prod_op}) into Fourier space:
\begin{eqnarray}
f\cdot g & = \int \tilde f(-\kk)g(\kk)\frac{\dd\kk}{(2\pi)^d},\\
\widetilde{Af}(\kk) & = \int \tilde A(\kk,-\kk')\tilde f(\kk')\frac{\dd\kk'}{(2\pi)^d},\\
\widetilde{AB}(\kk,\kk') & = \int \tilde A(\kk,-\kk'')B(\kk'',\kk')\frac{\dd\kk''}{(2\pi)^d}.
\end{eqnarray}
The Fourier transform of the translation-invariant operator $A(\xx,\xx')=a(\xx-\xx')$ reads
\begin{equation}
\tilde A(\kk,\kk')=(2\pi)^d\tilde a(\kk)\delta(\kk+\kk').
\end{equation}
Moreover, if $A$ is isotropic, its Fourier transform only depends on the norm $|\kk|$: $\tilde a(\kk)=\tilde a(|\kk|)$.

In this article, we do not use a different notation for the one-variable function associated to a translation-invariant operator: the number of variables indicates if we refer to the operator or to its associated function. For instance, we will use $\tilde \Delta(\kk,\kk')=(2\pi)^d\tilde \Delta(\kk)\delta(\kk+\kk')$.

\section*{References}

\bibliographystyle{plain_url}

\bibliography{biblio}

\end{document}